\documentstyle[11pt,newpasp_turku,twoside,epsfig]{article}
\def\edcomment#1{\iffalse\marginpar{\raggedright\sl#1\/}\else\relax\fi}
\reversemarginpar

\begin{document}
\title{Superluminal Motion and Relativistic Beaming in Blazar Jets}
 \author{K. I. Kellermann, M. L. Lister, and D. C. Homan}
\affil{NRAO, 520 Edgemont Rd., Charlottesville, VA, 22903, U.S.A.}
\author{E. Ros and J. A. Zensus}
\affil{MPIfR, Auf dem Hugel 69, Bonn, D-53121, Germany}
\author{M. H. Cohen and M. Russo}
\affil{Astronomy 105-24, Caltech, Pasadena, CA 91125, U.S.A.}
\author{R. C. Vermeulen}
\affil{ASTRON, P.O. Box 2, NL-7990 AA Dwingeloo, The Netherlands}

\begin{abstract}
High resolution radio observations remain the most direct way
to study the formation and evolution of radio jets associated with the
accretion onto massive black holes.  We report preliminary results of
our seven year VLBA observational program to understand the nature of
relativistic beaming in blazars and the surrounding
environment of massive black holes.

Most blazars show an apparent outward flow away from an active core.
However, in a few sources the motion appears inward, most likely
the result of projection of a curved trajectory which bends back toward
along the line of sight.  The apparent motion of jet features is not
always oriented along the direction separating the feature from the
core, and in a few cases we have observed a clear change in the
direction and velocity of a feature as it flows along the jet.  In
other sources, the motion appears to follow a simple ballistic
trajectory.  We find no simple relation between the time scales of
flux density changes and apparent component velocities.

\end{abstract}

\section{Background}

It is generally accepted that the blazar phenomenon is due to the
anisotropic boosting of the radiation along the direction of motion
which gives rise to an apparent enhanced luminosity at all wavelengths
if the observer is located close to the direction of motion.  There
are many observations which support this interpretation including the
one sided appearance of blazar jets and the rapid flux density
variability observed at many wavelengths. However, the only direct
observations of relativistic motion are at radio wavelengths when
motion close to the line of sight produces a compression of the time
frame resulting in apparent superluminal motion.  High resolution
interferometric radio images are able to measure such motions which
are typically less than one milliarcsecond per year.

Since 1994, we have been using the NRAO Very Long Baseline Array
(VLBA) at 15 GHz (2 cm) to study this relativistic outflow in a sample
of quasars and active galactic nuclei (AGN).  Our goal is to
understand the nature of the relativistic flow and the origin and
propagation of relativistic jets.  In particular, we want to know how
blazars differ from other quasars and active galactic nuclei.  The
high resolution radio images often show pronounced bends and twists.
We want to know whether or not the flow appears ballistic, that is if
individual features have straight trajectories as would occur from a
precessing nozzle,  or, whether features follow the curvature of the jet
characteristic of plasma instabilities.  Are there changes in the
speed or direction of features as they propagate down the jet?  Does
the moving pattern actually reflect the bulk flow velocity, or is
there a separate pattern velocity, for example reflecting the
propagation of shocks along the jet?  Is there a characteristic
Lorentz factor for different classes of AGN?  If not, what is the
distribution of Lorentz factors and what determines their value?  
Blazars typically show pronounced flux density
variations on time scales ranging from minutes to years.  Do these
flux density outbursts reflect the origin of new superluminal
components, and how do the time scales of intensity variations relate
to apparent velocity?

Our full sample consists of 173 galaxies, quasars, and BL Lac Objects.
In order to relate our observations to relativistic beaming models, we
wished to define a complete un-biased flux density limited sample.
However, as there are no sky surveys at 15 GHz, and AGN are generally
flux density variable, there is no simple objective way of obtaining a
precisely defined flux density limited sample.

Assuming a constant intrinsic value of the Lorentz factor, $\gamma$,
then if the bulk velocity is equal to the pattern velocity, it is easy
to calculate the distribution of observed apparent velocity (e.g.,
Vermeulen \& Cohen 1994).  However, if there is a distribution of
$\gamma$'s, then an analytic solution is more difficult.  One of our
goals was to compile a sample whose properties can be compared with
Monte Carlo simulations of relativistic beaming, so we selected
sources on the basis of the parsec scale flux density only,
ignoring any contribution from extended (kiloparsec-scale) structure
that is not necessarily beamed.  Our sample includes all known sources
which meet the following criteria.

\begin {itemize}

\item Declination $> -20^\circ$

\item Galactic latitude $|b| > 2.5^\circ$

\item Total 2 cm VLBA flux density $> 1.5$ Jy, ($>2$ Jy if below the
celestial equator) at any epoch since 1995.

\end {itemize}
We refer to this sample as an {\it unbiased representative sample}. We
have, so far, 
good multi epoch observations of 96 sources.  Observations in progress
are expected to increase the number of sources to what will be a complete 
unbiased sample
of about 120 sources.  We have constructed the
sample by reference to the K\"uhr 1 Jy catalog (K\"uhr 1981), the VLA
calibrator manual, the JVAS survey, the VLBA Calibrator Survey
(Beasley et al. 2002), the 22 GHz VLBI survey of Moellenbrock et
al. (1996) the high-frequency peaked samples of Ter{\"a}sranta et
al. (2001) and Dallacasa et al. (2000), and the UMRAO
(http://www.astro.lsa.umich.edu/obs/radiotel/umrao.html) and RATAN
(Kovalev et al. 1999) monitoring programs.  Although our selection
method is somewhat complex, it is based on the directly-measured compact
flux density, and does not use a single-epoch spectral index criterion
to estimate compact flux density. Also the fact that survey membership
is not determined from a single "snapshot" epoch means that we are
not excluding potentially interesting sources simply because they
happened to be in a low state at the time of the original
investigation.

In this paper we present data on the 157 individual features 
found in the 96 sources in our full sample which have 
core-jet structure for which we have obtained
good multi-epoch data on their motion and which have measured redshifts. 
We have generally observed each source
about once per year, but more often for those sources with rapid changes and
less often for those with slow changes.  Each image typically has an
angular resolution better than 1 milliarcsec, rms noise about 250
$\mu$Jy, and dynamic range better than 1000:1.  Images of all of our
observations are available at http://www.nrao.edu/2cmsurvey which will
soon be supplemented by material describing the motions observed in
each source.  Throughout this paper we use a cosmology with
$H_{0}=65$  km/sec/Mpc, $\Omega_{\Lambda}=0.7$, and $\Omega_{m}=0.3$.

\section{Statistics of Superluminal; Motion}

In spite of decades of studying superluminal source motions, the
details of the kinematics have remained elusive.  One of the problems
is, that contrary to indications of early observations (e.g., Cohen et
al. 1977), the radio jets often do not contain simple well defined
moving components.  Instead, the jets may show a complex brightness
distribution with regions of enhanced intensity that may brighten and
fade with time.  Some features appear to move; others are stationary,
or may break up into two or more separate features, and it is often
unclear how these moving features are related to the actual underlying
relativistic flow.

With these sensitive, high-resolution, high-dynamic range images from
the VLBA, we are generally able to define one or more components in
each source which have lasted for the duration of our observing
program.  Figure 1 shows the distribution of apparent linear velocity
for the 157 components contained in our full sample that have
well-determined  motions.  This includes 104 quasar components, 31 BL  Lac
components, and 22 components associated with the nucleus of an active
galaxy.

\begin{figure}[t]
\begin{center}
\epsfig{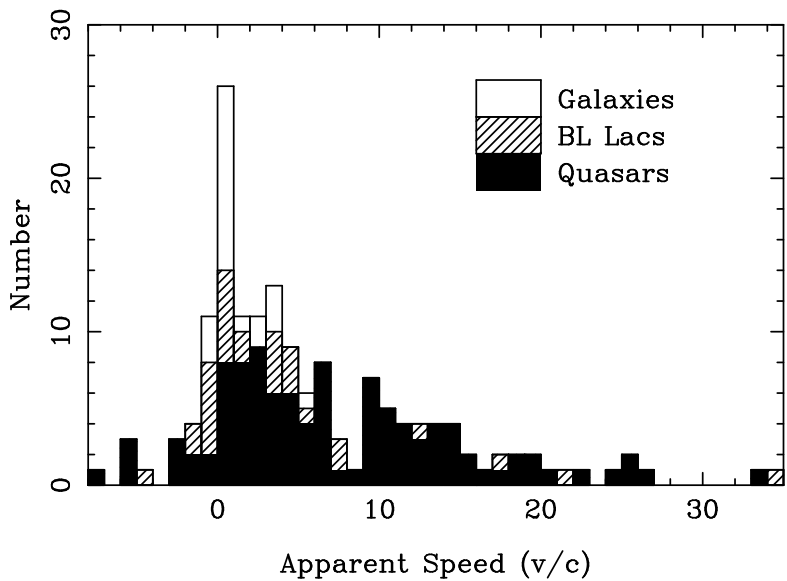}
\end{center}
\caption{Distribution of apparent linear velocity for the 156
individual features contained in our sample of 96 sources that 
have well determined motions.}  
\end{figure}

Figure 1 is in marked contrast to early discussions of superluminal
motion, which indicated typical values of $\gamma$ in the range 5 to
10 (Cohen et al. 1977, Porcas 1987).  We believe that these earlier
studies were biased in favor of faster apparent velocities since they
were not based on unbiased samples, but rather used velocities which
had been reported in the literature and which contained a
disproportionate fraction of high velocities, as the slow velocities
being "uninteresting" and more difficult to measure, were usually not
followed up with further observations nor were they reported in the
literature.

Most of the sources in our sample are quasars and their velocity
distribution is peaked near low values of {\it v/c} between zero and ten,
but there is a tail extending out to {\it v/c} $\sim$ 34. Features
associated with the active nuclei of galaxies all appear to have
motions in the range $0 < v/c < 8$, while the BL Lac objects appear
more uniformly distributed over the entire range from 0 to 35.

Several of the observed velocities are negative, that is the jet
component appears to be approaching rather than receding from the
core.  However, most of these reported negative velocities are
consistent, within the errors, with no significant motion.  In other
cases, there is evidence of a newly emerging component ejected from
the core, and the combination is not resolved by our beam. This
causes an apparent shift in the position of the core and a
corresponding decrease in the apparent separation of the core and jet
component.  It is also possible that the true core is not seen,
possibly due to absorption, and that both of the components we are
observing are part of the moving jet.  In a few cases, such as
0454+844, 0735+178, and 1128+385, the apparent decrease in component
separation from the core may be due to component motion away from the
core along a highly curved jet which bends back toward the line of
sight so that the apparent projected separation from the core appears
to decrease with time.

We interpret our observations based on relativistic beaming models
which assume all sources are relativistic with an intrinsic velocity
close to the speed of light described by a Lorentz factor, $\gamma$.
In a simple ballistic model, in which all jets have the same Lorentz
factor, the effect of Doppler boosting increases the probability of
observing sources close to the line of sight.  In the case of a flux
limited sample, angles close to $1/\gamma$ are commonly observed where
$\beta_{\mathrm app}$  $\sim \gamma$ (e.g., Vermeulen \& Cohen 1994, Vermeulen
1995).  If there is no Doppler boosting, then most sources are
expected to lie near the plane of the sky with an apparent velocity near
{\it c}.  The observed velocities do not show the expected concentration
near the upper end of the distribution corresponding to the simple
single-gamma ballistic model.  Ekers \& Laing (1990) have commented
that light echo models, in particular, which do not invoke any Doppler
boosting, are consistent with this kind of observed velocity
distribution.  Our observed distributions actually peak at
lower values than expected from simple light echo models.
Moreover, we find a strong correlation between apparent velocity and
apparent radio luminosity as expected if the apparent radio luminosity
is enhanced by Doppler boosting (Lister et al. 2003a).

Lister and Marscher (1997) have shown that an observed velocity
distributions similar to that shown in Figure 1, may be reproduced
with a power law distribution of intrinsic Lorentz factors.  Our data
are consistent with such a distribution having a large excess of small
Lorentz factors contained in a volume limited sample  (Lister et
al. in preparation).  Alternatively, the bulk velocity flow which
determines the amount of Doppler boosting may be less than the pattern
flow which may reflect the propagation of shock fronts rather than the
relativistic flow (e.g., Vermeulen $\&$ Cohen 1994).  The
interpretation of jet kinematics is further complicated if there is a
distribution of velocities within a single jet. For example, there may
be a fast inner jet surrounded by a more slowly moving outer
sheath.  In such cases the appearance and apparent velocity will be
a complex function of the orientation with respect to the line of
sight

\begin{figure}[t]
\begin{center}
\epsfig{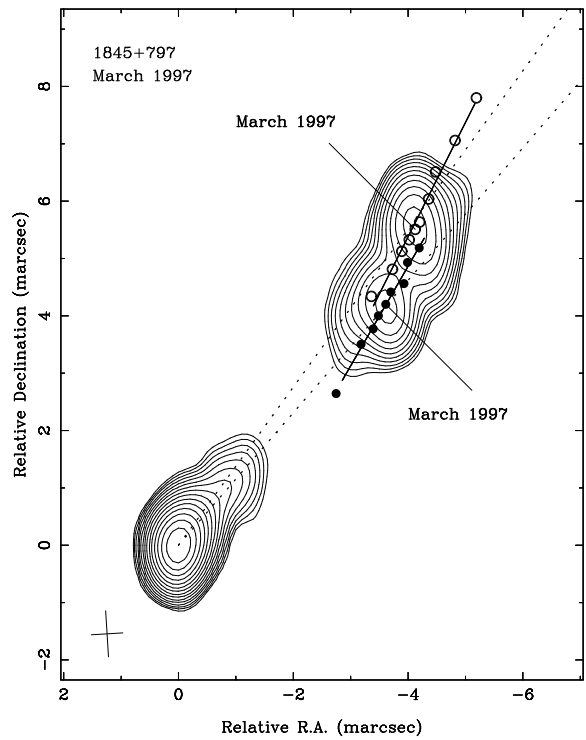}
\end{center}
\caption{2 cm VLBA image of the nucleus of the radio galaxy 3C 390.3
(1845+797) in March 1997 with the positions of the two
jet components superimposed for each of the epochs of our observations
from August 1994 to April 2001. The two dashed lines represent the
direction from the core toward the jet components in March 1997.}
\end{figure}

\section{Kinematics of Curved jets}

Many of the jets we have observed show pronounced curvature sometimes
with multiple oscillations characteristic of plasma instabilities.
Individual components may follow a wide range of trajectories.  In
some cases, such as 3C 273 (1226+023), the location of the bend appears
to propagate away from the center of activity.  In such cases the
motion may be described as ballistic, that is components appear to
move along a straight trajectory, but one which may not be pointed
toward the most compact feature assumed to be the core.  In other
cases, such as 3C 279 (1253$-$055), the component motion is more complex
and appears to flow along a non radial or even curved trajectory
(Homan et al. in preparation).

Figure 2 illustrates an example of non radial motion in the jet of the
quasar 3C390.3 (1845+797).  In this case the two prominent components
appear to be moving with similar velocities of $\sim 2.5c$, but along
slightly different trajectories of 27 and 30 degrees.  Neither are
aligned with the direction of the core or with the extended jet
feature which points toward the distant hot spot along position angle = 35\deg.

\begin{figure}[t]
\plotfiddle{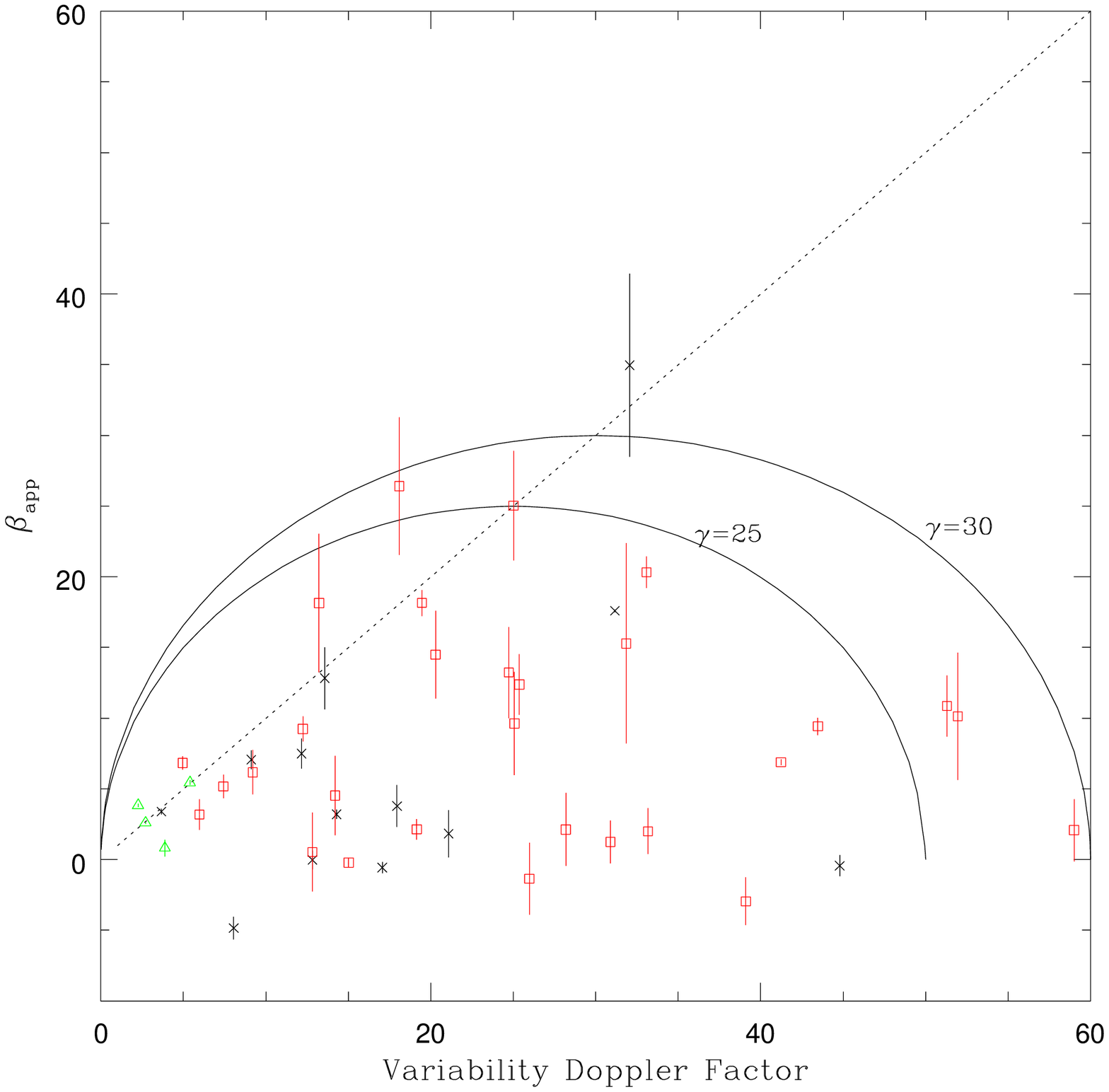}{3.5in}{0.}{45.}{45.}{-150}{-60}	
\caption{Apparent velocity, $\beta_{app} = {\it v/c}$ plotted against
Variability Doppler factor, $D_{var}$,
for the fastest component found in 49 sources calculated using the
method of L{\"a}hteenm{\"a}ki \& Valtaoja (1999).  The triangles refer to sources
in galaxies, the crosses to BL Lac objects, and the squares to
quasars. The curved lines shows the
expected locus of points for Lorentz factor values of 25 and 30.
Points below the diagonal line lie inside the $1/\gamma$ half width of
the beaming cone.}
\end{figure}
 
The Ultra Luminous Infra Red Galaxy 1345+125 (4C 12.50) which has a
very curved jet along with a weak counterjet, is consistent with a
conical helix of wavelength 280 pc that is the result of
Kelvin-Helmholtz instabilities driven by a slow precession of the jet
nozzle with a half-angle of 23 degrees oriented along an angle of 82
degrees to the line of sight (Lister et al. 2003b).
 
\section{Time Variability and Motions}

It is generally assumed that the large rapid changes in flux density
which are frequently observed in blazars are the result of Doppler
boosting combined with time compression of the relativistic plasma
moving nearly along the line of sight.  Small changes in the direction of 
the flow then can lead to large changes in the apparent luminosity
and velocity.  Provided that the bulk flow
velocity and the pattern velocity are the same, we would expect to see a
relation between variability time scale and the apparent transverse
linear velocity.  L{\"a}hteenm{\"a}ki $\&$ Valtaoja (1999) have calculated
Doppler boosting factors from Mets{\"a}hovi 1.3 cm and 8 mm variability
data assuming an intrinsic brightness temperature of $5\times  10^{10}$\,K
characteristic of a self absorbed synchrotron source where the
particle and magnetic energy is in equilibrium.

We have compared our VLBA observations with both the Mets{\"a}hovi and
2 cm University of Michigan (UMRAO) variability data. The UMRAO data
covers a longer time period and is more densely sampled; however
events which appear to be well defined at 8 mm and 1.3 cm are often
blended at the longer wavelength.  Figure 3 shows that the
observations essentially all fit inside the $\gamma = 30$ curve, as
they should for a flux density limited sample (Lister $\&$ Marsher
1997), but the detailed distribution does not match the expected one.
Further, there are a number of highly variable sources sources, such
as 3C 454.3, CTA 102, and 2134+00 which show little or no significant
motions.  Each of these sources has a highly bent or oscillating jet.
Possibly, the bright features which we have observed in these sources
reflect stationary positions along the jet where the relativistic flow
is oriented close to the line of sight thus giving rise to enhanced
radiation due to Doppler boosting.  There does appear to be a well
defined upper limit to the measured apparent velocity which is close
to the variability Doppler factor. Calculations of the variability
Doppler factor using intrinsic brightness temperatures closer to the
inverse Compton limit lead to even larger scatter and the lack of any
clear envelope to the distribution.  Also, we note that there is a
large dispersion between between Doppler factors deduced from the
UMRAO 2 cm data and the shorter wavelength Mets{\"a}hovi data, so the
robustness of the Doppler factors calculated in this way appears to be
very uncertain.  The relation between $\beta_{\mathrm app}$ and
$D_{\mathrm var}$ will be discussed in more detail by Cohen et al. (in
preparation).

\bigskip

We acknowledge valuable discussions with Hugh and Margo Aller, Yuri
Kovalev jr. and Matthias Kadler.

\end{document}